\documentclass[12pt]{article}
\usepackage{graphicx}
\usepackage{times}

\newenvironment{natabstract}{%
\begin{quote} \bf}
{\end{quote}}

\title{Frequency-stabilization to 6$\times$10$^{-16}$ via spectral-hole burning}

\author
{Michael J. Thorpe,$^{1\ast}$ Lars Rippe,$^{2}$ Tara M. Fortier$^{1}$, Matthew S. Kirchner$^{1}$,\\and Till Rosenband$^{1}$
\\
\normalsize{$^{1}$National Institute of Standards and Technology,}\\
\normalsize{325 Broadway St., Boulder, CO 80304, USA}\\
\normalsize{$^{2}$Department of Physics, Atomic Physics, Lund University,}\\
\normalsize{Professorsgata 1, 223 63 Lund, Sweden}\\
\\
\normalsize{$^\ast$Address correspondence to; E-mail:  mthorpe@nist.gov.}
}

\date{\today}

\begin{document}

\maketitle

\newcommand{\EuYSO}{Eu$^{3+}$:Y$_2$SiO$_5$~}
\newcommand{\Eu}{Eu$^{3+}~$}
\newcommand{\Xition}{$^7$F$_0$ $\rightarrow$ $^5$D$_0$ }
\newcommand{\YSO}{Y$_2$SiO$_5$ }

\begin{natabstract}

  We demonstrate two-stage laser stabilization based on a combination of
  Fabry-P\'erot and spectral-hole burning techniques.  The laser is first
  pre-stabilized by the Fabry-P\'erot cavity to a fractional-frequency
  stability of $\sigma_y(\tau) < 10^{-13}$.  A pattern of spectral holes
  written in the absorption spectrum of \EuYSO serves to further stabilize
  the laser to $\sigma_y(\tau) = 6\times10^{-16}$ for $2~$s$~\leq\tau\leq 8~$s.
  Measurements characterizing the frequency sensitivity of \EuYSO spectral holes
  to environmental perturbations suggest that they can be more frequency-stable
  than Fabry-P\'erot cavities.
\end{natabstract}

\section{Introduction}

Frequency-stable laser-local-oscillators (LLOs) are limiting
components of the new generation of optical atomic clocks.  Quieter
LLOs would allow optical clocks to run more stably, yielding faster
comparison measurements among different clocks, for more precise
gravitational measurements and tests of fundamental physics
\cite{Rosenband2008,Blatt2008,ChouSci2010}. Quieter LLOs may also
yield lower-phase-noise microwave oscillators by frequency-division
via femtosecond laser frequency combs \cite{Bartels2005,Fortier2011}.

\indent State-of-the-art LLOs are based on lasers that are tightly locked
to high-finesse Fabry-P\'{e}rot (FP) cavities constructed from low thermal
expansion glass.  The Pound-Drever-Hall (PDH) locking technique \cite{Drever1983}
provides sufficient signal-to-noise ratio (SNR) to lock the laser to the FP cavity,
such that the laser's fractional frequency stability beyond $\tau$ = 0.1~s is limited
only by the fractional optical-length stability of the cavity. Over the past decade,
the stability of these systems has improved from $3\times10^{-16}$ for durations of
1 to 100 seconds \cite{Young1999}, to a recent demonstration of $2\times10^{-16}$ for
durations of 2 to 10 seconds~\cite{Jiang2011}. During this time, practical
concerns such as vibration-sensitivity have been addressed
\cite{Notcutt2005,Nazarova2006,Webster2007,Thorpe2010}, but lowering
the noise floor has proved challenging. Numata~\textit{et al.}
identified thermo-mechanical noise as the fundamental physical
effect that limits the length stability of such cavities \cite{Numata2004},
and this insight was confirmed experimentally \cite{Notcutt2006}.
Thermo-mechanical noise may be reduced by using longer cavities~\cite{Jiang2011},
by choosing cavity, mirror and mirror-coating materials with a reduced
mechanical loss tangent, by operating the cavity at lower temperatures
or by using larger-diameter optical modes.  However, all of these approaches
present significant technical difficulties.

\indent Spectral-hole burning in cryogenically cooled crystals has been
demonstrated as an alternative to FP cavities for laser-stabilization
\cite{Sellin1999,Strickland2000,Bottger2003,Julsgaard2007}.  In such systems,
the instantaneous laser frequency excursions are compared to the spectral memory
stored in a crystal, and an error signal is derived to stabilize the laser.
A fractional-frequency stability of $\sigma_y(\tau)=3\times10^{-14}$ for
$\tau=10$ ms has been demonstrated \cite{Strickland2000}, but at longer
times the instability was significantly higher due to the transient nature of
holes in Tm$^{3+}$:Y$_2$Al$_5$O$_{12}$ and high photon fluxes that acted to
degrade the spectral memory.  Although initial demonstrations of spectral-hole-burning
laser locks did not reach the stability of FP-cavities, we expect the fundamental
thermomechanical noise limit to be lower, because spectral holes are atomic frequency
references that are perturbed only weakly through coupling to the crystal host.  Hence,
the spectral-hole reference frequency is largely decoupled from the crystal's
thermomechanical noise.

\indent Several properties of \EuYSO have motivated high-resolution spectroscopy
of spectral holes.  Both the ground ($^7F_0$) and excited ($^5D_0$)
states of the \Eu laser stabilization transition have small magnetic moments,
making them insensitive to magnetic field fluctuations \cite{Shelby1981}.  The
$^5D_0$ excited state has a lifetime of 1.9~ms and photon echo measurements
indicate linewidths as low as 122~Hz for the \Xition transition
\cite{Yano1991, Equall1994}.  When \Eu is doped into the \YSO crystal, each
ion experiences a slightly different crystal field that causes a different Stark
shift for each absorber. As a result, the \EuYSO absorption spectrum is inhomogeneously
broadened to a width that depends linearly on the doping concentration
\cite{Konz2003,Sellars2004}.  For the present work, we used a doping of 0.5\% Eu$^{3+}$,
corresponding to an inhomogeneous linewidth of 10~GHz that, in principle, can contain
nearly 10$^8$ resolvable spectral holes. \EuYSO spectral holes can also be long-lived.
At T = 2.0~K, a lifetime of 20 days has been measured \cite{Konz2003}.  However, the
lifetime depends strongly on temperature.  For our nominal operating temperature
(T = 4.5~K) the lifetime is more than 1 day, but the spectral holes last only 1~s at
15~K.  Therefore, a spectral-hole pattern can be used for long-term laser stabilization
at T~$\leq$~4.5~K, and can be easily erased if the crystal temperature is temporarily
increased.

\begin{figure}[t!]
\centering
\includegraphics[width=4in]{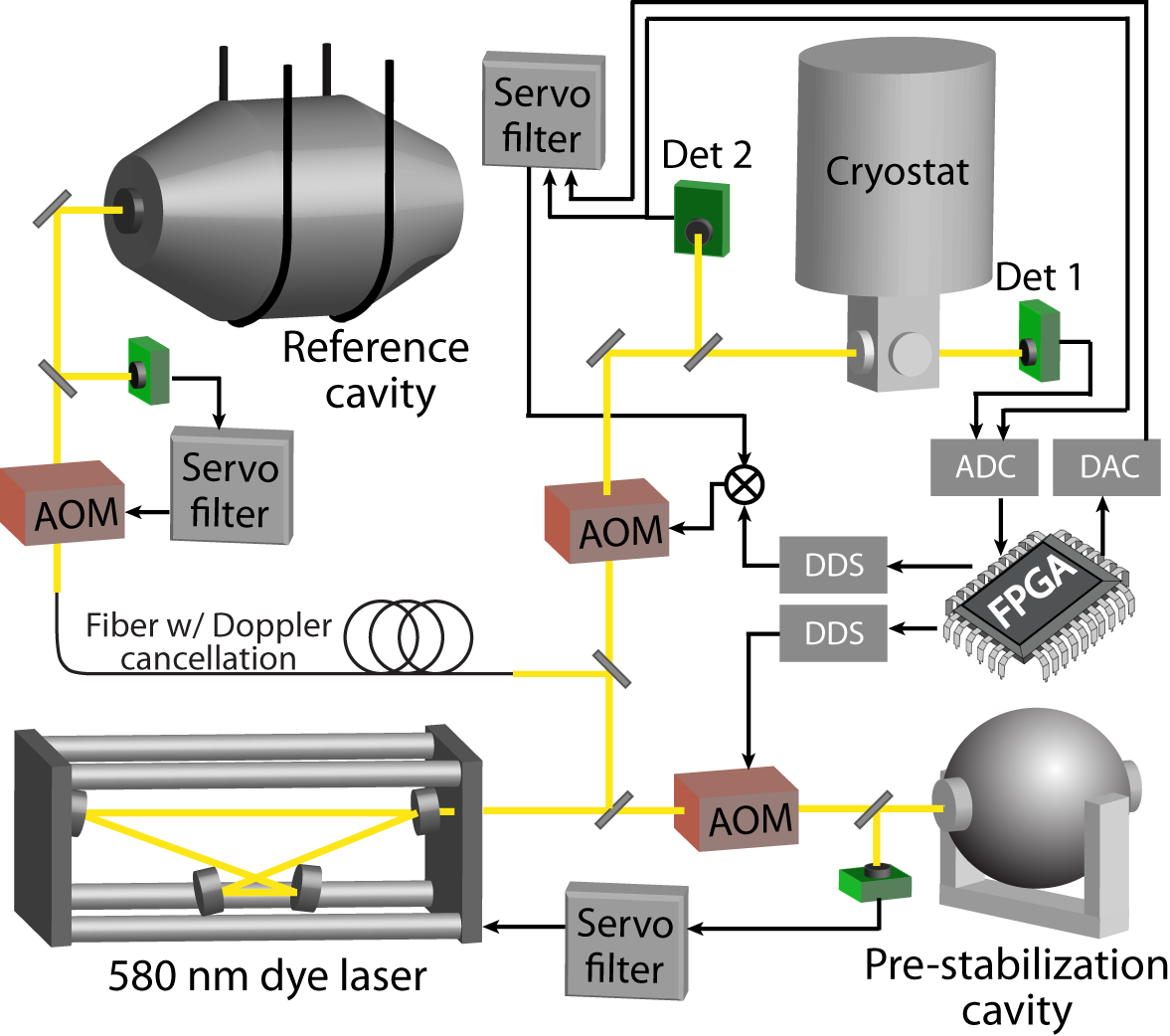}
\caption{The \EuYSO spectroscopy and laser stabilization experimental setup.
Spectroscopy is performed with a pre-stabilized 580~nm dye laser that
illuminates a \EuYSO crystal housed in an optical flow cryostat.  A field-programmable
gate array (FPGA) with an embedded microprocessor controls the details of the spectroscopy
including laser frequency, light intensity and pulse durations.  The frequencies of spectral
holes can be compared against the frequency of the pre-stabilization cavity, or against
a reference cavity that typically stabilizes the Al$^+$ clock laser.} \label{Fig1}
\end{figure}

\indent Here we describe a two-stage laser-stabilization system (Fig.~\ref{Fig1}), where
a FP cavity first stabilizes the laser frequency to $\sigma_y(\tau) \approx 10^{-14}$
($0.1~s < \tau<10~s)$.  The pre-stabilized laser is then modulated to address $10^{15}$ atomic
absorbers within a 5~mm length and 1~cm diameter \EuYSO crystal.  This servo simultaneously
writes and stabilizes the laser to a pattern of spectral holes in the \EuYSO absorption spectrum.
By using a pre-stabilized laser and many spectral holes we derive an error signal for laser
stabilization while reducing perturbations to the spectral memory.  Here we demonstrate the
use of 10 to 100 spectral holes to stabilize a 580~nm laser to $\sigma_y(\tau) = 6 \times 10^{-16}$
for $2~s<\tau<8~s$. We also report several properties of \EuYSO that make this material a promising
candidate for achieving higher laser stability than is currently available via optical cavities.

\section{Results}

\begin{figure}
\centering
\includegraphics[width=5in]{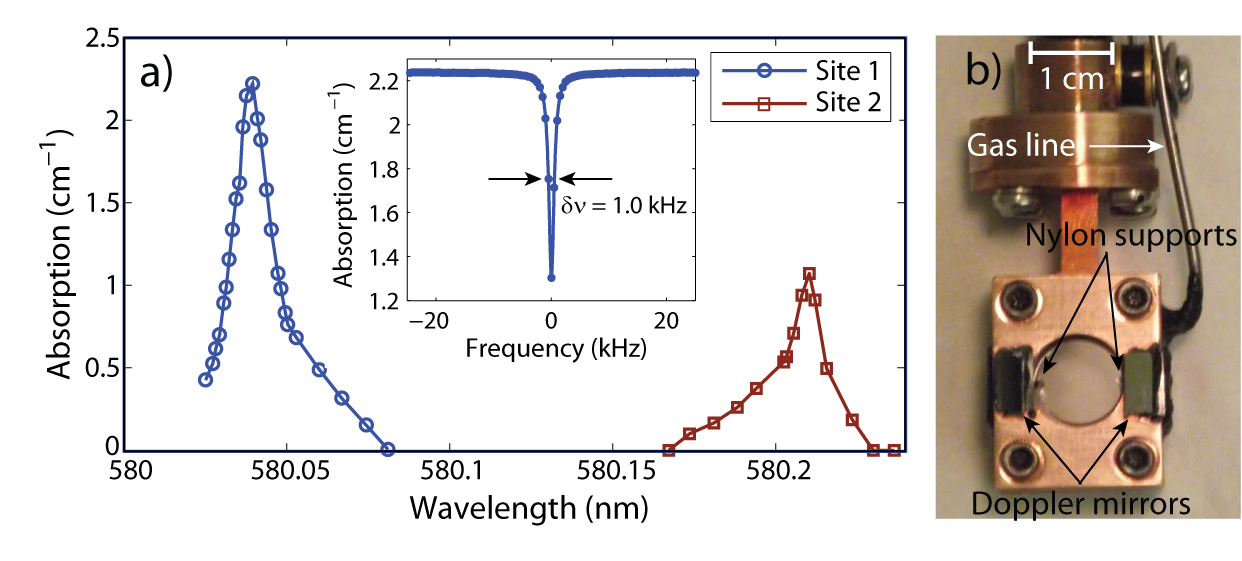}
\caption{a) The absorption spectrum of the \Xition transition in 0.5\% doped \EuYSO at
T = 4.5~K.  The two inhomogeneously broadened absorption features `sites 1 \& 2' result from
the two different positions within the \YSO unit cell where \Eu can substitute for Y$^{3+}$.
The inset shows a single spectral hole written at 580.0390 nm and T = 4.5~K with a burn time of
100~ms and a laser intensity of 35~$\mu$W/cm$^2$.  This hole was measured with 2~ms probe pulses, each
separated by 500~Hz (SNR = 1500). b) A picture of the sealed chamber that provides a controlled
pressure environment and acceleration insensitive mounting for the \EuYSO crystal.} \label{Fig2}
\end{figure}

\indent The absorption spectrum for the \Xition transition in 0.5\% doped \EuYSO at T = 4.5~K is
shown in figure~\ref{Fig2}(a).  The two resolved peaks at vacuum wavelengths of 580.0390~nm and
580.2110~nm result from two different locations within the \YSO crystal unit cell where \Eu can
substitute for Y$^{3+}$.  These features are referred to as sites 1 and 2, respectively.  The
inset in figure~\ref{Fig2}(a) shows a single spectral hole with a linewidth of 1.0~kHz, a 40\%
contrast and an SNR of 1500, written at 580.0390~nm and T = 4.5~K.  Although photon-echo measurements
suggest that \EuYSO spectral holes as narrow as 100~Hz are possible \cite{Equall1994}, so far we
have not observed linewidths narrower than 500~Hz.  This broader linewidth is partially due to phonon
scattering that contributes 300~Hz to the homogeneous linewidth at T~=~4.5~K \cite{Konz2003}.
Further broadening may be caused by vibrations or fluctuations in electric or magnetic fields in our
cryostat, which can be improved in future systems.  We used two protocols for writing and
detecting spectral holes.  For characterizing the sensitivity to environmental perturbations we
wrote the holes at a relatively high intensity (35~$\mu$W/cm$^2$ to 140~$\mu$W/cm$^2$) for durations
of 10~ms to 200~ms, and detected the holes at lower intensities (3~$\mu$W/cm$^2$) with shorter pulse
durations (1~ms).  For laser stabilization we wrote and detected the pattern with the low intensity
(3~$\mu$W/cm$^2$) and short duration (1~ms) pulses.

\begin{figure}[t!]
\centering
\includegraphics[width=3.5in]{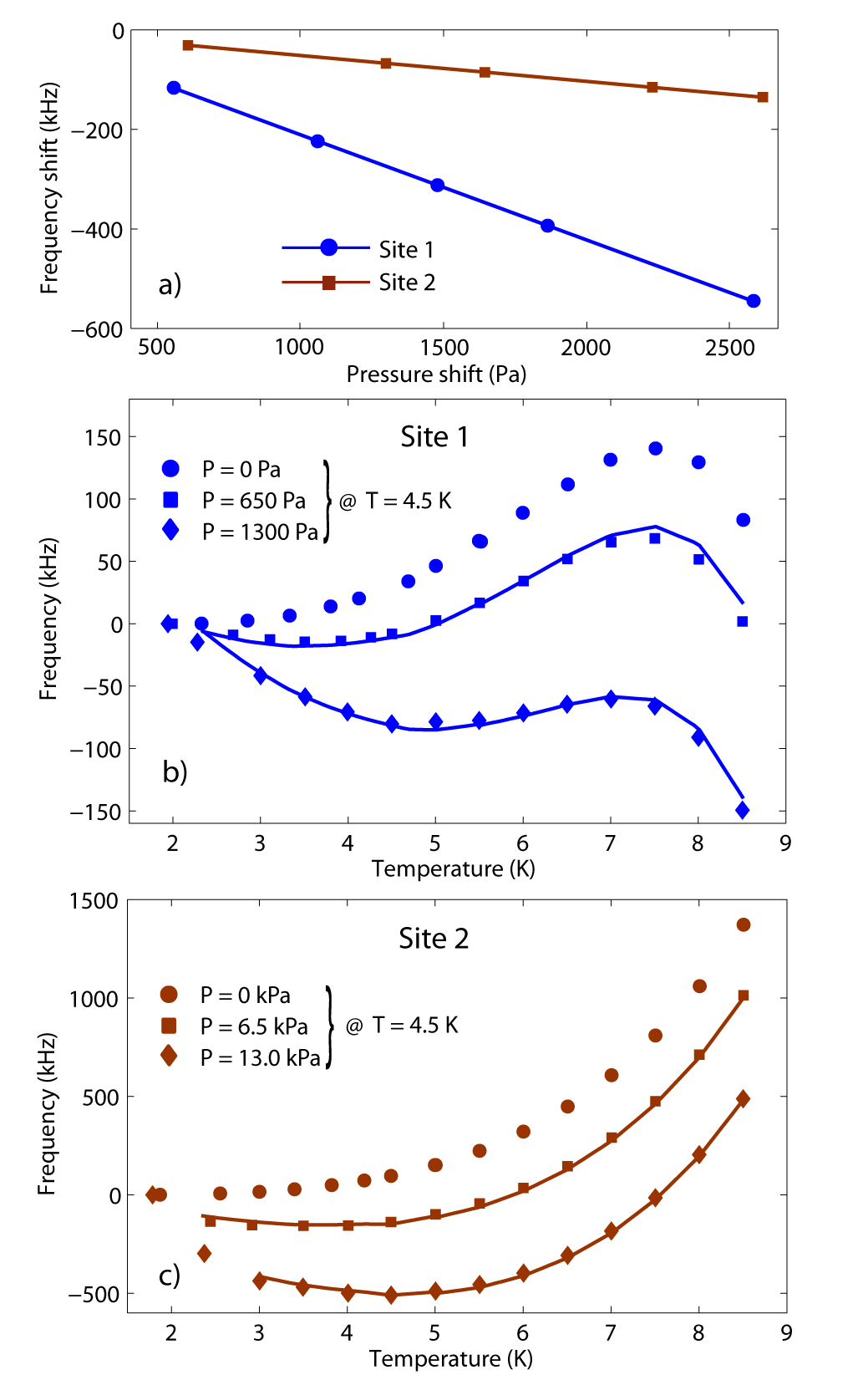}
\caption{The pressure and temperature sensitivity of the spectral-hole frequency
for sites 1 and 2.  a) Pressure sensitivity is $\alpha_1$ = -211.4(4)~Hz/Pa for holes
at site 1 and $\alpha_2$ = -52.0(7)~Hz/Pa at site 2.  b) \& c)  The temperature sensitivities
for holes at sites 1 \& 2 are given by the P~=~0~Pa data (circles).  In a sealed chamber,
the gas pressure and temperature are linked by the ideal gas law, and the pressure and temperature
sensitivities both contribute to the frequency of the spectral hole.  At certain combinations
of P and T for each site, the frequency becomes first-order insensitive to the temperature of
its environment.  The solid lines through the P$\neq$0 data are expected frequency shifts given
the measured pressure and temperature sensitivities and the best fit volume ratio (see text).}
\label{Fig3}
\end{figure}

\indent The \EuYSO crystal is housed in an optical flow cryostat that provides the ability to
continuously vary the crystal temperature from 2.0~K to 300~K (Fig.~\ref{Fig1}). Inside the
cryostat, the crystal is enclosed in a second sealed chamber that is filled with helium gas,
so that the pressure and temperature environment can be controlled independently (Fig.~\ref{Fig2}(b)).
A gas line leads from the sealed chamber to a manifold outside the cryostat where the pressure is
controlled and monitored.  Figure~\ref{Fig3} shows the pressure and temperature sensitivity of the
spectral-hole frequency.  The pressure sensitivities, $\alpha_1$ = -211.4(4)~Hz/Pa for site 1
and $\alpha_2$ = -52.0(7)~Hz/Pa for site 2, are shown in figure~\ref{Fig3}(a).  By combining
the pressure sensitivity with the bulk modulus of \YSO (135 GPa), we calculate the sensitivity
of spectral holes to changes in the volume of the crystal ($\delta$f/f~=~0.055$\times\delta$V/V for
site 1 and $\delta$f/f~=~0.014$\times\delta$V/V for site 2). Compared to FP cavities where
$\delta$f/f~=~$0.33\delta$V/V and the typical bulk modulus is 34~GPa~\cite{CorningULE},
spectral holes have significant isolation from mechanical instability that not only makes
them less sensitive to accelerations, but also reduces their susceptibility to thermal noise.

\indent Measurements of the temperature sensitivity for sites 1 and 2 are shown in
figures~\ref{Fig3}(b) \& \ref{Fig3}(c) respectively.  The data show the temperature sensitivities
$\kappa_i$ for sites ($i\in \{1,2\}$) in the absence of a background gas (P~=~0~Pa). Site 1 displays
a lower temperature sensitivity than site 2, and an anomalous backward bending of the curve at 7.5~K.
These curves provide a high-resolution extension of data presented by K\"onz \emph{et al.}
\cite{Konz2003}.  When the sealed chamber is filled with helium gas, the pressure and temperature of
the gas are related through the ideal gas law.  Furthermore, since the pressure sensitivity has a
negative slope, and the temperature sensitivity has a positive slope (for 2.0~K$<T<$7.5~K) certain
combinations of pressure and temperature lead to a vanishing first-order term of the temperature
sensitivity.  This effect is illustrated in figures~\ref{Fig3}(b) \& \ref{Fig3}(c).  Because the gas
manifold is outside the cryostat, at room temperature, the gas behaves as a two-reservoir system.  The
equation of state for this system (see Methods section) can be combined with the
pressure sensitivity $\alpha_i\Delta P$ and the zero-pressure temperature sensitivity $\kappa_i(T_c)$
functions for each site to determine the spectral-hole temperature sensitivity in the presence of
helium gas:
\begin{equation}
f_i(P,T_c) = \kappa_i(T_c) + \alpha_iP(T_c).
\label{Sensitivity}
\end{equation}
Here $f_i$ is the spectral-hole frequency, $P(T_c)$ is the equation of state and $T_c$ is the temperature
of the crystal inside the cryostat.  A fit to the measured data determines the ratio of the cryogenic to
room temperature gas volumes for the current system, $V_c/V_r$ = 0.022(3). This ratio is important because
it determines how strongly temperature and pressure fluctuations in the gas manifold (outside the cryostat)
will affect the temperature and pressure of the cryogenic gas volume. The temperature-insensitive points for
spectral holes have some advantages over the analogous `zero-crossing temperature' of the  coefficient of
thermal expansion for ultra-low expansion glass (ULE) cavities.  First, the temperature of the insensitive
point for spectral holes can be tuned by changing the pressure of helium gas in the sealed chamber.  Second,
the spectral holes exhibit reduced temperature sensitivity compared to ULE.  For ULE, the quadratic temperature
sensitivity is 720~kHz/K$^2$ for a cavity operating at 580~nm~\cite{CorningULE}.  The values for \EuYSO
spectral-holes are 16~kHz/K$^2$ for site 1 and 114~kHz/K$^2$ for site 2.

\begin{figure}[b!]
\centering
\includegraphics[width=4.5in]{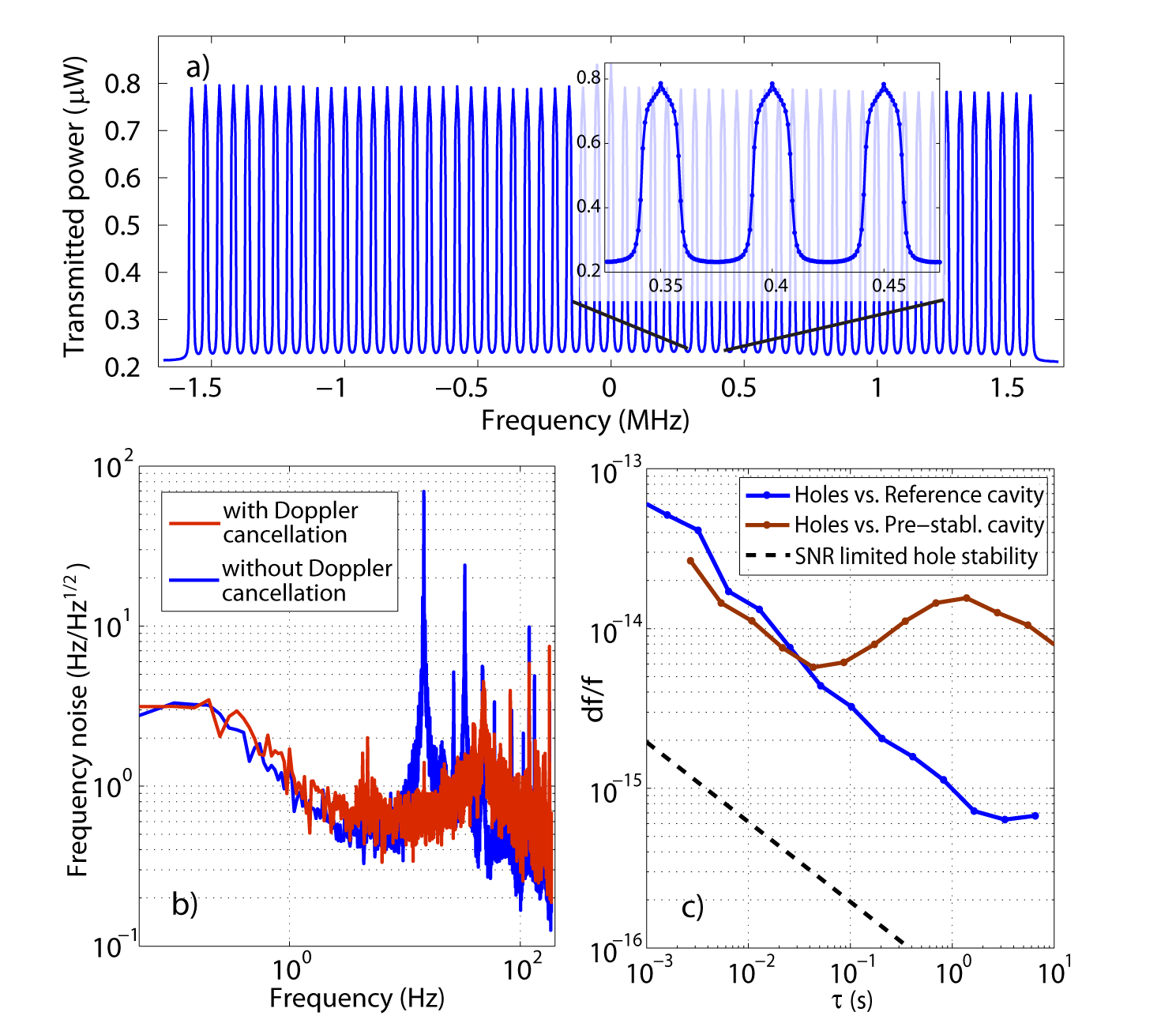}
\caption{a) A pattern of 61 spectral holes spaced by 50~kHz for laser stabilization.  The inset
shows three of the holes in higher resolution. b) Laser frequency noise spectra of the spectral-hole laser lock with and without Doppler cancellation of the crystal motion (30 mHz resolution bandwidth). The motional resonance at 14.5~Hz sets an upper limit on the spectral-hole acceleration sensitivity of 7$\times$10$^{-12}$g$^{-1}$. c) Allan deviation traces of frequency comparisons between the pre-stabilization cavity, the reference cavity, and the spectral-hole pattern. The pre-stabilized laser is made artificially noisy, $\sigma_y(\tau) > 10^{-14}$ (red trace), and is used to write and lock to a pattern a spectral holes.  The spectral-hole lock is then compared to the frequency of a reference cavity that has a noise floor of $\sigma_y(\tau) \approx 5\times$10$^{-16}$ (blue trace).  The black dashed line is the expected short-term stability based on the spectral-hole linewidth and the SNR of the absorption measurements.} \label{Fig4}
\end{figure}

\indent A pattern of spectral holes, simultaneously written and used for laser stabilization, is shown
in figure~\ref{Fig4}(a).  For details about the writing and probing of the spectral-hole pattern, refer
to the Methods section.
During laser stabilization, the pressure and temperature of the crystal environment are held
at an insensitive point.  At the same time, a Doppler measurement cancels frequency shifts in the
spectral-hole pattern due to the motion of the crystal~\cite{JCB1994}.  This measurement is made
with a Michelson interferometer where one arm of the interferometer is formed by a mirror attached
to the crystal chamber (see Fig.~\ref{Fig2}(b)).  Figure~\ref{Fig4}(b) shows the frequency spectrum
of the spectral-hole-stabilized laser with and without Doppler cancellation.  Two prominent resonances
in the crystal motion appear at 14.5~Hz and 33~Hz, corresponding to RMS frequency shifts of 12.1~Hz
and 4.3~Hz respectively.  When Doppler cancellation is implemented, these shifts are reduced, improving
the frequency stability of the hole pattern.  The residual frequency shift of 0.25~Hz amplitude (RMS)
due to the 14.5~Hz motion provides an estimate of the acceleration sensitivity of the spectral-hole
pattern of 7$\times$10$^{-12}$~g$^{-1}$ (1~g = 9.8~m/s$^2$), which is an order of magnitude below the
lowest passive acceleration-sensitivity of FP cavities~\cite{Jiang2011,Webster2007,Leibrandt2011}.
As in FP cavities, the acceleration sensitivity depends on the mounting configuration, and further
improvements are possible.  For the present work, the crystal is supported on two 0.8~mm thick and
2.5~mm wide nylon tabs~(Fig.~\ref{Fig2}(b).  The support points are near the vertical center of the
crystal such that the frequency of spectral holes are nominally insensitive to
accelerations~\cite{Notcutt2005}.

\indent  To determine the frequency stability of the spectral-hole laser lock, we independently locked
the 580~nm laser to a reference cavity that is typically used to stabilize the clock laser for the
$^{27}$Al$^+$ optical clock~\cite{Chou2010AlAl}.  To ensure that the write/probe laser was locked tightly
to the spectral-hole pattern, we injected noise into the pre-stabilization servo, thereby degrading
the laser stability to $\sigma_y\approx10^{-14}$.  Figure~\ref{Fig4}(c) shows the write/probe laser stability
and the frequency comparison between the reference cavity and the spectral-hole laser lock  with the linear
frequency drift subtracted. The reference cavity has a noise floor of $\sigma_y(\tau) \approx 5\times$10$^{-16}$,
and a typical drift rate of 0.1~Hz/s. During the comparison measurements, we observed relative drifts between
the spectral-hole pattern and the reference cavity that varied between 0.1~Hz/s to 0.5~Hz/s.  A noise floor
of $\sigma_y = 6.5\times10^{-16}$ was observed in the comparison.   So far, we have not observed
signal-to-noise-ratio-limited short-term stability of the spectral-hole laser lock (see Fig.~\ref{Fig4}(c)).
We expect that this is due to a combination of pressure fluctuations of the gas reservoir, vibrations in
cryostat, and Dick-effect noise from the write/probe laser due to the 90$\%$ duty cycle of our probe pulses.
Each of these sources of instability must be addressed to achieve SNR-limited performance.

\section{Discussion}
\indent Our initial attempts at laser stabilization have already demonstrated performance
that is competitive with the best FP cavities on short time scales.  Further improvements are
expected from the use of two probe beams that enable interleaved measurements of the hole
pattern to eliminate Dick effect noise.  To reach a stability of $\sigma_y(\tau) = 3\times$10$^{-17}/\sqrt\tau$
will require $\Delta P < 67~\mu$Pa and $\Delta a < 4~\mu$g at 1~s of averaging time, and can
be achieved through better engineering of the cryogenic environment.  Magnetic field
sensitivity measurements indicate that a modest control of the magnetic field $\Delta B < 10^{-5}$~T
will be required. Finally, due to the pressure and temperature cancellation at the site 1 insensitive
point, an overall temperature stability of $\Delta T < 1.4$~mK  is sufficient, but temperature gradients
must also be minimized.  This technique for achieving low temperature sensitivity may be useful for other
proposed frequency references based on solids such as Thorium-doped crystals \cite{Peik2003,Rellergert2010}.

\indent At longer time scales, frequency comparison measurements are currently limited by the lifetime
of the spectral-hole pattern.  This limitation needs to be addressed if spectral-hole burning
laser stabilization is to become useful for most LLO applications.  One approach for dealing with
this problem is to write a self-regenerating pattern that eventually achieves a quasi-steady
state.  Such a spectral-hole pattern is possible if the pattern of holes is wide enough to
encompass all of the \Eu hyperfine ground states, and the hole spacing is small enough
to prevent population buildup between adjacent spectral holes.  Stabilization to such a
steady-state pattern must also compensate for the side-holes and anti-holes that accompany
each spectral feature~\cite{Yano1991}.

\indent If the technical sources of frequency noise can be made sufficiently small, the stability
of \EuYSO spectral holes will be determined by thermal nosie.  Our measurements of the pressure sensitivity
provide some insight into this fundamental limit.  If we assume an isotropic crystal geometry and a
conservative mechanical loss angle for \EuYSO ($\phi = 10^{-3}$) then thermal noise for a crystal of
the size used in the present work implies a fractional-frequency instability below
3$\times$10$^{-17}$~\cite{Numata2004}.  However, many crystalline materials exhibit reduced mechanical
loss at cryogenic temperatures, therefore the mechanical loss angle at 4.5~K may be smaller
($\phi \approx 10^{-5}$) leading to a thermal noise floor that is an order of magnitude
lower.  Nevertheless, \YSO has low crystal symmetry, and a full anisotropic treatment of its thermo-mechanical
properties is required to make accurate estimates of the thermal noise floor~\cite{Heinert2010}, as
well as to engineer optimal strategies for mounting these crystals to minimize environmental sensitivity.

\indent During the past decade, the frequency-stabilities of LLOs based on either FP cavities or spectral-hole
burning have improved by 30\%, and new optical clocks have a substantial need for further improvements.  A
combination of the two laser-stabilization approaches has the potential to yield order-of-magnitude stability
gains due to the low sensitivity of \EuYSO spectral holes to environmental perturbations and internal noise.
Such gains could benefit a wide range of LLO applications including the realization of atomic time, gravitational
measurements for geodesy, radar and communications applications and very long baseline interferometry.

\section{Methods}

\indent The experimental setup for high-resolution spectroscopy of \EuYSO spectral
holes is shown in Fig.~\ref{Fig1}.  The output of a 580~nm dye laser is split into
three beam-lines that provide pre-stabilization of the laser frequency, frequency
comparisons with a reference cavity, and a write/probe beam for \EuYSO spectroscopy.
The laser is locked to the pre-stabilization cavity using the PDH scheme.  With
optimal locking, the laser can achieve a stability of $\sigma_y(\tau) = 1.2\times10^{-15}$
($0.5~s<\tau<12~s$) \cite{Leibrandt2011}. For spectral-hole laser locking experiments we
intentionally reduce the stability of the write/probe laser
($\sigma_y(\tau) > 10^{-14}$ for $0.3~s<\tau<6~s$) to clearly demonstrate the stability of
the spectral holes.  An acousto-optic modulator (AOM) is placed before the cavity, enabling
laser frequency adjustments of the `reference cavity' and `write/probe' beam-lines that are
independent of the pre-stabilization cavity lock.

\indent For temperatures above 4.2~K, the crystal temperature is controlled by a servo that
heats helium vapor surrounding the sealed chamber. For temperatures below 4.2~K, the heater
is turned off and a scroll pump and a pressure regulator are used to lower the pressure, and
hence the temperature, of the helium vapor.  In the range from 2.0~K to 10.0~K the temperature
can be controlled to within 1~mK.  The sealed chamber is filled with helium gas by a 0.9~mm
inner diameter tube that extends from the chamber inside the cryostat to the room temperature
gas manifold.  A capacitive pressure gauge located on the room temperature gas manifold is used
to monitor the pressure inside the sealed chamber.  The pressure and temperature of this system
are related by
\begin{equation}
P = \frac{k_BCT_c}{1+\frac{T_cV_r}{T_rV_c}},
\label{PVT}
\end{equation}
where $P$ is the pressure of the system, $k_B$ is the Boltzmann constant, $T_r$ ($T_c$) correspond to
the room (cryostat) temperature, $V_r$ ($V_c$) are the room temperature (cryogenic) gas volumes and C
is the total number of helium atoms divided by $V_c$.

\indent \EuYSO spectroscopy is performed by measuring the frequency-dependent absorption of light
transmitted through the \EuYSO crystal.  For the current work, the crystal was illuminated by a
6~mm diameter beam.  The intensity, frequency, and duration of the illumination are controlled
by a microprocessor that is embedded in a field programmable gate array (FPGA).  The FPGA instructs a
digital to analog converter (DAC) to output voltage pulses to an intensity servo to control the optical
power and pulse duration. Prior to each pulse, the FPGA sets the frequency of a direct digital synthesizer
(DDS) that drives the broadband AOM located before the cryostat.  In the current setup, the frequency of
the incident beam can be tuned over 600~MHz of bandwidth, and the beam intensity can be set to values
between 0.1~$\mu$W/cm$^2$ and 200~$\mu$W/cm$^2$.  The incident and transmitted beams, detected by `Det 1'
and `Det 2' and digitized by an analog-to-digital converter (ADC), are used to calculate the absorption
for each experiment.

\indent For laser stabilization experiments, the embedded microprocessor controls the process for
simultaneously writing and probing the spectral-hole pattern.  Experimental parameters such as the
number of holes, hole spacing, center frequency of the pattern, servo gains, pulse duration and
pulse power are adjustable by the user.  To generate a spectral-hole pattern, a single hole
is first burned with successive pulses until a target hole-depth (typically 30\% contrast)
is reached.  Each time a hole reaches the target depth, the burning process for a new hole
is initiated until the pattern contains the pre-specified number of holes.   When holes exceed
the target depth they are deemed ready for laser stabilization.  During the burning process there
are two categories of holes, those ready for stabilization and those still in the burning process.
The servo interleaves burning pulses on unfinished holes with frequency stabilization measurements
on sufficiently deep holes until all holes have exceeded the target depth.  For each frequency
measurement and burn operation, a random hole is selected from the corresponding group. This
bootstrapping process establishes a regular pattern of spectral holes.

\indent Frequency stabilization measurements are made by recording the absorption on both sides of a
hole at roughly 2/3 of the maximum depth.  These measurements are subtracted, multiplied by the
servo gain, and fed-back to the frequency of the AOM positioned before the pre-stabilization cavity.
As the holes are continually measured, they become both deeper and wider.  The frequency measurement
data is used in a secondary feedback loop to track the width of each hole to ensure that servo
measurements are always made at roughly 2/3 of the hole depth. For each frequency measurement
the hole is also burned at the center frequency until its contrast approaches 100\% so that each
hole retains a single-peaked shape. Once all holes reach full contrast, only frequency measurements
are performed and the measurement duty cycle is 90$\%$. Finally, a small amount of gain is used to
adjust the laser probe frequency for each hole to remove frequency errors that occur because the
holes are written with a noisy laser.

\indent To determine the stability of the pre-stabilization cavity and spectral-hole
pattern laser locks, frequency comparisons are made with a reference cavity.  Light from
the dye laser is delivered to the reference cavity over a Doppler-noise-canceled fiber.
The laser is locked to the reference cavity with a PDH error signal is fed-back to an
AOM positioned before the cavity.  When locked, the AOM drive frequency provides a measure
of the frequency difference between the reference cavity and the pre-stabilization cavity
(if the spectral-hole laser lock is inactive) or the spectral-hole laser lock when it is
active.

\section{Acknowledgements}
M. J. Thorpe acknowledges support from the National Research Council.  The authors thank R. L.
Cone, J. C. Bergquist, J. Ye, J. L. Hall and D. J. Wineland for useful discussions, and
D. R. Leibrandt and J. A. Sherman for help with manuscript preparation.  This work is
supported by DARPA and ONR and is not subject to U.S. copyright.

\section{Author Contributions}
M.J.T, T.R. and L.R. designed the experiments. M.J.T., T.M.F. and M.S.K. performed the experiments.
M.J.T. and T.R. conducted the data analysis.  M.J.T., T.R., L.R. and M.S.K. wrote the manuscript.


\begin{thebibliography}{10}

\bibitem{Rosenband2008}
T.~Rosenband, {\it et~al.\/}, {\it Science\/} {\bf 28}, 1808-1812 (2008).

\bibitem{Blatt2008}
S.~Blatt, {\it et~al.\/}, {\it Phys. Rev. Lett.\/} {\bf 100}, 140801-1-4 (2008).

\bibitem{ChouSci2010}
C.~W. Chou, D.~B. Hume, T.~Rosenband, D.~J. Wineland, {\it Science\/} {\bf
  329}, 1630 (2010).

\bibitem{Bartels2005}
A.~Bartels, S.~A.~Diddams, C.~W.~Oates, G.~Wilpers, J.~C.~Bergquist, W.~H.~Oskay, L.~Hollberg, {\it Opt. Lett.\/} {\bf 30}, 667-669 (2005).

\bibitem{Fortier2011}
T.~M. Fortier, {\it et~al.\/}, {\it arXiv:1101.3616v2\/}  (2011).

\bibitem{Drever1983}
R.~W.~P. Drever, J.~L.~Hall, F.~V.~Kowalski, J.~Hough, G.~M.~Ford, A.~J.~Munley, H.~Ward, {\it App. Phys. B\/} {\bf 31}, 97-105 (1983).

\bibitem{Young1999}
B.~C. Young, F.~C. Cruz, W.~M. Itano, J.~C. Bergquist, {\it Phys. Rev. Lett.\/}
  {\bf 82}, 3799-3802 (1999).

\bibitem{Jiang2011}
Y.~Y. Jiang, A.~D.~Ludlow N.~D.~Lemke, R.~W.~Fox, J.~A.~Sherman, L.-S.~Ma, C.~W.~Oats, {\it Nat. Photonics\/} {\bf 5}, 158-161 (2011).

\bibitem{Notcutt2005}
M.~Notcutt, L.~S. Ma, J.~Ye, J.~L. Hall, {\it Opt. Lett.\/} {\bf 30}, 1815-1817
  (2005).

\bibitem{Nazarova2006}
T.~Nazarova, F.~Riehle, U.~Sterr, {\it Appl. Phys. B\/} {\bf 83}, 531-536 (2006).

\bibitem{Webster2007}
S.~A. Webster, M.~Oxborrow, P.~Gill, {\it Phys. Rev. A\/} {\bf 75}, 011801-1-6
  (2007).

\bibitem{Thorpe2010}
M.~J. Thorpe, D.~R. Leibrandt, T.~M. Fortier, T.~Rosenband, {\it Opt.
  Express\/} {\bf 18}, 18744-18751 (2010).

\bibitem{Numata2004}
K.~Numata, A.~Kemery, J.~Camp, {\it Phys. Rev. Lett.\/} {\bf 93}, 250602-1-4
  (2004).

\bibitem{Notcutt2006}
M.~Notcutt, L.-S.~Ma, A.~D.~Ludlow, S.~A.~Foreman, J.~Ye, J.~L.~Hall, {\it Phys. Rev. A\/} {\bf 73}, 031804-1-4 (2006).

\bibitem{Sellin1999}
P.~B. Sellin, S.~N. M., J.~L. Carlsten, R.~L. Cone, {\it Opt. Lett.\/} {\bf
  24}, 1038-1040 (1999).

\bibitem{Strickland2000}
N.~M. Strickland, P.~B. Sellin, Y.~Sun, J.~L. Carlsten, R.~L. Cone, {\it Phys.
  Rev. B\/} {\bf 62}, 1473-1476 (2000).

\bibitem{Bottger2003}
T.~B\"ottger, G.~J. Pryde, R.~L. Cone, {\it Opt. Lett.\/} {\bf 28}, 200-202 (2003).

\bibitem{Julsgaard2007}
B.~Julsgaard, A.~Walther, S.~Kr\"oll, L.~Rippe, {\it Opt. Express\/} {\bf 15},
  11444-11465 (2007).

\bibitem{Shelby1981}
R.~M. Shelby, R.~M. Macfarlane, {\it Phys. Rev. Lett.\/} {\bf 47}, 1172-1175 (1981).

\bibitem{Yano1991}
R.~Yano, M.~Mitsunaga, N.~Uesugi, {\it Opt. Lett.\/} {\bf 16}, 1884-1886 (1991).

\bibitem{Equall1994}
Equall, R.~W., Y.~Sun, R.~L. Cone, R.~M. Macfarlane, {\it Phys. Rev. Lett.\/}
  {\bf 72}, 2179-2182 (1994).

\bibitem{Konz2003}
F.~K\"onz, Y.~Sun, C.~W.~Thiel, R.~L.~Cone, R.~L.~Equall, R.~L.~Htchenson, R.~M.~Macfarlane, {\it Phys. Rev. B\/} {\bf 68}, 085109-1-9 (2003).

\bibitem{Sellars2004}
M.~J. Sellars, E.~Fraval, J.~J. Longdell, {\it J. Lumin.\/} {\bf 107}, 150-154
  (2004).

\bibitem{CorningULE}
Corning ule datasheet,
  http://www.corning.com/docs/specialtymaterials/pisheets/ulebro91106.pdf.

\bibitem{JCB1994}
J.~C. Bergquist, W.~M. Itano, D.~J. Wineland, {\it International School of
  Physics ``Enrico Fermi"\/}, T.~W. H\"ansch, M.~Inguscio, eds. (North-Holland,
  Amsterdam, 1994).

\bibitem{Leibrandt2011}
D.~R. Leibrandt, M.~J.~Thorpe, M.~Notcutt, R.~E.~Drullinger, T.~Rosenband, J.~C.~Bergquist, {\it Opt. Express\/} {\bf 19}, 3471-3482  (2011).

\bibitem{Chou2010AlAl}
C.~W. Chou, D.~B. Hume, J.~C.~J. Koelemeij, D.~J. Wineland, T.~Rosenband, {\it
  Phys. Rev. Lett.\/} {\bf 104}, 070802-1-4 (2010).

\bibitem{Peik2003}
E. Peik, Chr.~Tamm, {\it Europhys. Lett.\/} {\bf 61}, 181-186 (2003).

\bibitem{Rellergert2010}
W.~G. Rellergert, D.~DeMille, R.~R.~Greco, M.~P.~Hehlen, J.~R.~Torgerson, E.~R.~Hudson, {\it Phys. Rev. Lett.\/} {\bf 104}, 200802-1-4
  (2010).

\bibitem{Heinert2010}
D.~Heinert, {\it et~al.\/}, {\it Journal of Physics: Conerference Series\/}
  {\bf 228}, 012032-1-6 (2010).

\end{thebibliography}
\end{document}